\documentclass[10pt,journal,compsoc]{IEEEtran}
\ifCLASSOPTIONcompsoc
  \usepackage[nocompress]{cite}
\else
  \usepackage{cite}
\fi

\ifCLASSINFOpdf
\else
\fi

\usepackage{amsmath,amsfonts}
\usepackage[linesnumbered,lined,boxed,commentsnumbered]{algorithm2e}

\usepackage{algorithmic}
\usepackage{algorithm}
\usepackage{array}
\usepackage[caption=false,font=normalsize,labelfont=sf,textfont=sf]{subfig}
\usepackage{textcomp}
\usepackage{stfloats}
\usepackage{url}
\usepackage{verbatim}
\usepackage{graphicx}
\usepackage{cite}
\usepackage{tabularx}
\usepackage{xcolor}
\usepackage[normalem]{ulem}

\begin{document}
\title{FEDQ-Trust: Efficient Data-Driven Trust Prediction for Mobile Edge-Based IoT Systems}
\author{Jiahui Bai, Hai Dong,~\IEEEmembership{Senior Member,~IEEE,} Athman Bouguettaya,~\IEEEmembership{Fellow,~IEEE}
\IEEEcompsocitemizethanks{
\IEEEcompsocthanksitem Jiahui Bai and Hai Dong are with the School of Computing Technologies, RMIT University, Melbourne, VIC 3000, Australia.
Emails: s3735049@student.rmit.edu.au; hai.dong@rmit.edu.au
\IEEEcompsocthanksitem Athman Bougettaya is with the School of Computer Science, The University of Sydney, Darlington, NSW 2008, Australia.
E-mail: athman.bouguettaya@sydney.edu.au}%
\thanks{Manuscript received February 01,  2024; revised.}}

\markboth{IEEE Transactions on Services Computing,~Vol.~xx, No.~xx, February~2024}%
{Shell \MakeLowercase{\textit{et al.}}: Bare Advanced Demo of IEEEtran.cls for IEEE Computer Society Journals}

\IEEEtitleabstractindextext{%
\begin{abstract}
We introduce FEDQ-Trust, an innovative data-driven trust prediction approach designed for mobile edge-based Internet of Things (IoT) environments. The decentralized nature of mobile edge environments introduces challenges due to variations in data distribution, impacting the accuracy and training efficiency of existing distributed data-driven trust prediction models.
FEDQ-Trust effectively tackles the statistical heterogeneity challenges by integrating \underline{F}ederated \underline{E}xpectation-Maximization  with \underline{D}eep \underline{Q} Networks.
Federated Expectation-Maximization's robust handling of statistical heterogeneity significantly enhances trust prediction accuracy.
Meanwhile, Deep Q Networks streamlines the model training process, efficiently reducing the number of training clients while maintaining model performance.
We conducted a suite of experiments within simulated MEC-based IoT settings by leveraging two real-world IoT datasets. 
The experimental results demonstrate that our model achieved a significant convergence time reduction of 97\% to 99\% while ensuring a notable improvement of 8\% to 14\% in accuracy compared to state-of-the-art models. 
\end{abstract}

}

\maketitle

\IEEEdisplaynontitleabstractindextext

%
\IEEEpeerreviewmaketitle

\ifCLASSOPTIONcompsoc
\IEEEraisesectionheading{\section{Introduction}\label{sec:introduction}}
\else
\section{Introduction}
\label{sec:introduction}
\fi
\IEEEPARstart{T}{rust}  of an Internet of Things (IoT) service is a measure of how much belief a service consumer has in the service provider's ability to deliver the requested IoT service \cite{4146807}.
IoT services encompass offerings that leverage cloud computing and IoT technology, facilitating the interconnection of diverse devices, sensors, and systems via the Internet. This enables the collection, storage, processing, and analysis of data while offering a range of applications and services. \cite{7785887}.
Trust serves as a crucial prerequisite for ensuring the safety and reliability of IoT services \cite{Yan2014ASO}. 
IoT devices can potentially deliver subpar or even malicious services when they lack a measure of trustworthiness. This poses significant risks to consumers.
For example, trust emerges as a crucial factor in the scenario of sensing-based vehicle traffic navigation for the selection of sensing service providers (e.g. autonomous vehicles) \cite{cheng2021general}.
In this context, a vehicle's ability to make safe and precise decisions hinges on its trust in information accuracy and timeliness, sourced from multiple service providers. The vehicle prioritizes providers offering the most trustworthy information for critical road maneuvers, including lane changes and speed adjustments.
Therefore, it is imperative to recognize the significance of trust in IoT services and accord it the utmost attention.

Existing trust prediction approaches can be broadly categorized into model-driven and data-driven. Model-driven methods rely heavily on domain experts to define and interpret trust characteristics within specific application contexts \cite{7505635} \cite{Yan2014ASO}\cite{8246999}. Data-driven trust prediction utilises machine learning and statistical analysis to discern and quantify trust relationships within IoT services \cite{9583862}. Data-driven approaches offer a more dynamic and adaptive framework in comparison to model-driven approaches. They can more efficiently process large volumes of data from various sources, including service consumers' feedback, QoS evaluations, and real-time service performance metrics \cite{8329991}. This stream of methods allows for a more comprehensive and nuanced understanding of trust attributes like availability, reliability, and risk. Consequently, data-driven trust prediction aligns well with the complex and ever-evolving nature of IoT ecosystems, where the reliability and accuracy of services are paramount. 
IoT service providers can anticipate and address potential trust issues more effectively by leveraging these advanced analytical techniques. They can a higher level of service quality and consumer satisfaction.
Therefore, the adoption of data-driven trust prediction is not only a strategic choice but a necessity in the context of IoT trust management \cite{9583862}.

Mobile Edge Computing (MEC) represents an innovative computing paradigm that decentralizes computational resources, relocating them closer to data sources and end-users. This strategic placement significantly boosts data transmission and processing efficiency. MEC is proven to be particularly valuable when dealing with substantial trust information generated for IoT services. It provisions a unique capability for cellular base stations to deliver computing and storage resources to nearby IoT devices \cite{Beck2014MobileEC} \cite{hu2015mobile}. This relocation of resources to the network's edge addresses shortcomings inherent in traditional network designs. These shortcomings primarily include increased latency and network congestion due to the centralized processing of high-volume IoT data \cite{Beck2014MobileEC}\cite{hu2015mobile}. These shortcomings can significantly hinder the performance and reliability of IoT systems \cite{7786106}\cite{BOTTA2016684}. 
Such bottlenecks are effectively mitigated by distributing computing tasks across the network with the adoption of MEC. Therefore, MEC can ensure faster and more efficient data handling. In this regard, MEC effectively alleviates these bottlenecks. Furthermore, MEC's architectural design enables the provisioning of low-latency services, dramatically reducing data transmission times across the network. This characteristic is of paramount importance in real-time applications where delays could adversely impact system performance and user experience.

\begin{figure*}[htbp]
    \centering
    \includegraphics[width=18cm]{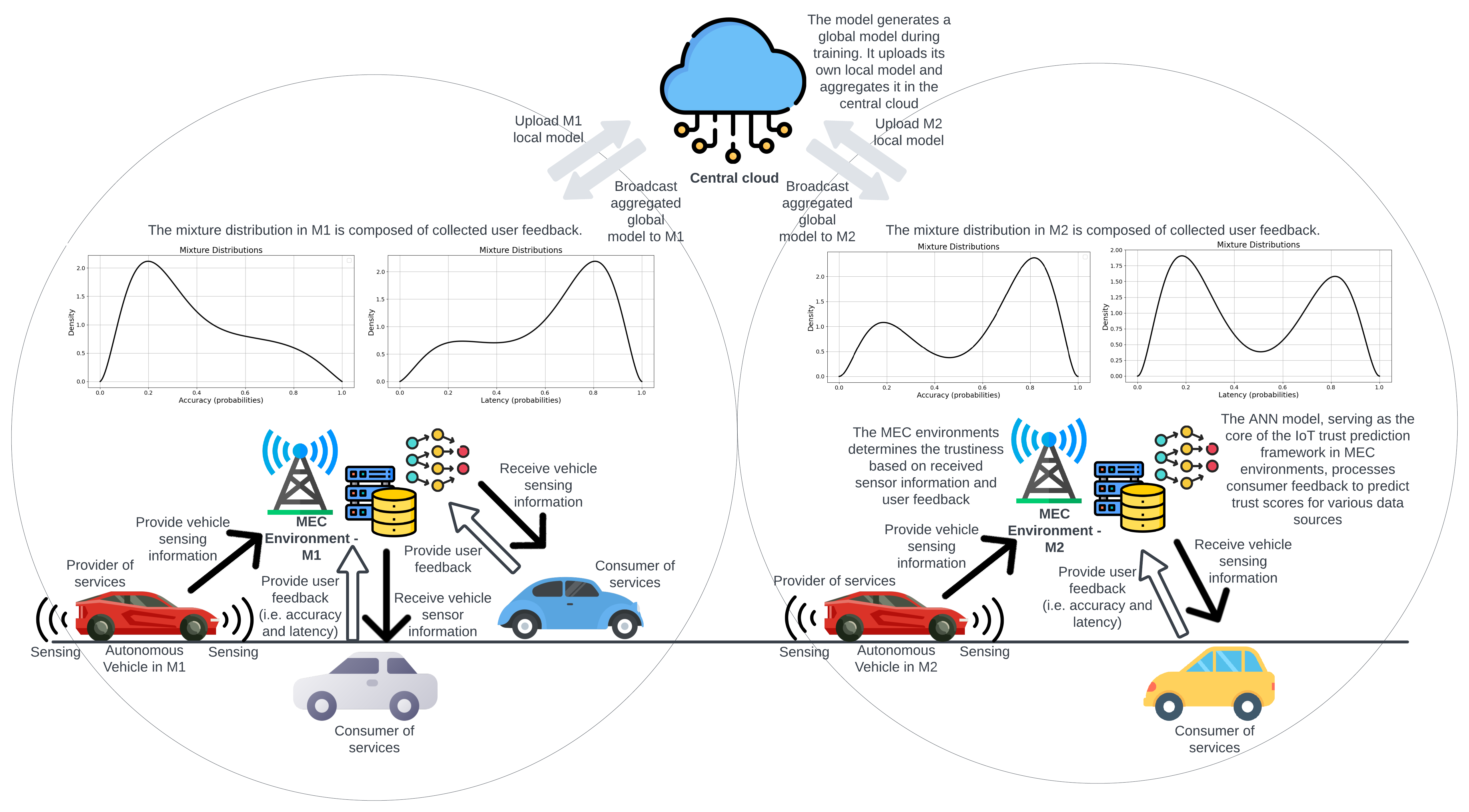}
    \caption{
    Schematic of trust prediction in MEC environments for autonomous vehicle-based sensing services.
    Autonomous vehicles (e.g., the red car) provide sensing services to aid the traffic navigation of other vehicles in multiple MEC environments. 
    Trust plays a critical role in selecting reliable sensing services. User feedback (e.g., user-perceived accuracy and latency) is collected from service consumers in each MEC environment. Variations in vehicle traffic, and user perception create various mixture distributions of trust data in different MEC environments. Edge devices in each MEC environment employ local machine learning models for sensing service trust prediction based on the trust data. A global model is formed in the central cloud by aggregating the parameters of local models. The parameters of the global model are then shared with each MEC environment to enhance the generalization capabilities of trust prediction. The mixture distributions of trust data challenge existing prediction models following this paradigm.
    }
    \label{fig:autonomous}
\end{figure*}

However, relying solely on MEC infrastructure is insufficient to fully address trust prediction challenges for IoT services. 
In such a context, trust information is typically accumulated in a massively distributed manner within each MEC environment, creating distributed data islands. 
The data-driven trust prediction models need to enable training and prediction in these environments by using collaborative strategies or federated learning \cite{9583862} \cite{8818406} \cite{10147216}.
These models aim to share local model parameters across different MEC environments to create a global model. 
The global model aids in wider generalization within different MEC environments. It integrates information from multiple diverse data sources to capture more comprehensive domain knowledge. Thus, it enables more accurate and reliable predictions across different MEC environments.
Each MEC environment generates trust information of varying quantities and particular distributions, leading to statistical heterogeneity in the form of non-identical and independent data distribution (non-IID) \cite{9583862}. Usually, these particular distributions are regarded as stemming from a mixture of underlying distributions \cite{smith2017federated}. Existing models often assume data follows a specific distribution pattern. In reality, the distribution of service data varies by location, time, and usage patterns. This introduces biases and errors in these models when predicting under different distributions, leading to inaccurate trust assessments in various MEC environments.
Therefore, the data-driven trust prediction model needs to overcome the \textit{statistical heterogeneity} for the same service provided in different MEC environments.

Figure \ref{fig:autonomous} demonstrates a motivation scenario of IoT service trust prediction in MEC environments. 
Autonomous vehicles (e.g., the red car) provide sensing services to aid the traffic navigation of other vehicles in multiple MEC environments. However, services with low trust may furnish inaccurate traffic information, adversely affecting safety and efficiency. Therefore, 
the prediction of service trust becomes essential to address this, particularly through the analysis of consumers' feedback data. The system gains insights into the trustworthiness of different services by gauging user-perceived accuracy and latency of traffic prediction.
MEC devices 
are leveraged to collect the feedback data and perform trust prediction in each MEC environment (i.e. M1 and M2), considering the large service area and the need for low latency. 
Each MEC environment trains a local model using the collected data.
They upload the local models to the central cloud.
The central cloud aggregates these local models 
to create a global model. The global model offers generalization capabilities across different MEC environments. The global model is shared among the MEC environments for local training.
The differences in the number of vehicles and traffic conditions as well as the variations in user perception may lead to distinct mixture distributions of trust data in M1 and M2. The mixture distributions challenge existing distributed data-driven prediction models that are mostly based on federated learning. These models are often designed to adapt to more homogeneous data patterns. They fail to capture underlying distribution differences across various MEC environments. This may result in low-accuracy trust prediction.
Data-driven IoT service trust prediction in MEC environments faces another critical technical challenge, i.e., the resource constraints of MEC devices.
Trust prediction needs to be conducted efficiently under these resource constraints to meet the demands of rapid service selection decision-making and delivery. While accuracy remains a crucial requirement for trust prediction, it is not feasible to simply increase the number of training rounds to achieve higher accuracy, given the resource limitations of IoT devices. More training rounds may lead to increased computational and communication resource consumption, which may be unsustainable for IoT devices \cite{9155494}. Therefore, an effective balance must be struck between high accuracy and low training overhead to address the reality of \textit{resource-constrained IoT devices}.

The features of statistical heterogeneity and resource intensity within the MEC-based IoT environment introduce a multitude of challenges to the existing data-driven IoT service trust prediction approaches as follows.
\begin{itemize}
\item[1)] 
Existing data-driven trust prediction methods (e.g., \cite{9583862}, \cite{8818406} and \cite{10147216}) have failed to adequately address the heterogeneity of IoT service trust information in MEC-based environments. They fall short in addressing the complex challenges posed by statistical heterogeneity, particularly in mixture distributions. Most of the existing research has been centred on improving the accuracy of trust models in MEC environments. However, they have not adequately explored the heterogeneity of IoT service trust information in these settings. This research gap underscores the necessity for a more specialized approach. This approach is required to effectively address the unique challenges posed by data distribution and sample size variations across different MEC settings.
\item[2)]
Currently, there are no solutions that can achieve sufficient training efficiency while meeting high accuracy requirements. Existing IoT service trust prediction models, as outlined in studies like \cite{9583862}, \cite{8818406}, \cite{10147216} and \cite{10248268}, fail to simultaneously meet the requirements of high accuracy and low training overhead. This limitation is clearly demonstrated in our experiments. Additionally, there is a noticeable lack of in-depth analysis regarding model convergence within the context of MEC environments.
\end{itemize}

This paper describes a data-driven IoT service trust prediction approach, named \underline{F}ederated \underline{E}xpectation-Maximization with \underline{D}eep \underline{Q}-Networks (FEDQ-Trust). This approach enables accurate trust prediction for IoT services in statistically heterogeneous MEC environments with higher training speeds. We summarize our specific contributions addressing the aforementioned challenges as follows.

\begin{itemize}
\item[1)] 
We frame the problem of IoT service trust prediction in MEC environments as a federated optimization problem to tackle the issue of statistical heterogeneity. The federated optimization problem is characterized by a mixture of diverse data distributions, specifically, a mixture distribution of IoT trust information. Our objective is to optimize a global prediction model capable of accommodating the various data distributions present in different MEC environments. To achieve this, we employ a Federated Expectation-Maximization (FedEM) framework \cite{marfoq2021federated}. 
FedEM effectively addresses the data imbalance problem across different MEC environments, thus resolving the statistical heterogeneity issue. This framework identifies shared implicit features among the underlying data features across diverse environments. It promotes collaboration and improves model accuracy while accelerating convergence.
\item[2)] 
We introduce a novel Deep Q Network (DQN)-based reinforcement learning approach to meet the demand for efficient model training. This method is designed to intelligently select a subset of MEC environments for federated optimization. The DQN-based approach optimizes the selection process by choosing a relatively small number of MEC environments, minimizing the computational and communication overhead. This results in a significant reduction in model convergence time, a critical factor when dealing with resource-constrained IoT devices. Despite the reduction in the number of participating MEC environments, our method ensures that model accuracy is not compromised. This is achieved through the intelligent, data-driven selection process facilitated by DQN. The selected subset continues to contribute effectively to the learning and predictive capabilities of the global model by identifying the most representative environments.
\end{itemize}

We conducted a thorough comparison with state-of-the-art distributed data-driven IoT trust prediction techniques \cite{9583862} \cite{10147216} to validate the effectiveness of our proposed methods. We utilized two IoT datasets to ensure a rigorous and unbiased evaluation. The datasets provide a practical and authentic context for our assessments. Our methods, especially FEDQ-Trust30, exhibited significant advancements over existing approaches. Across both datasets, FEDQ-Trust30 demonstrated a notable improvement of 8\% to 14\% in accuracy compared to the baseline models. More impressively, it achieved a significant reduction of 97\% to 99\% in elapsed time for model training. These results underscore the exceptional balance our methods offer between accuracy and convergence speed, significantly enhancing both aspects in comparison to current models in diverse IoT environments.

The structure of the remaining sections in this paper is as follows: Section 2 provides a review of prior research that forms the foundation for our work. 
Section 3 provides essential background and preliminary information for our study, offering readers the necessary context.
In Section 4, we formally define the problem through mathematical expressions, establishing a clear problem statement for our paper. 
Section 5 delves into the comprehensive explanation of our proposed solution. 
Subsequently, Section 6 offers detailed insights into the experiments conducted to evaluate the effectiveness of our proposed solution. 
Finally, in Section 7, we bring our work to a conclusion and engage in a discussion regarding potential directions for future research.

\section{Related works}
In this section, we mainly assess the prior research that appears before our proposed work in three main categories: 1) model-driven IoT trust prediction, 2) centralized data-driven IoT trust prediction and 3) distributed data-driven IoT trust prediction. We highlight the major shortcomings in these previous studies in Section 2.1, Section 2.2 and Section 2.3, respectively.

\subsection{Model-Drive IoT Service Trust Prediction}
In the realm of model-driven IoT trust prediction, \cite{wang2019crowdsourcing} introduces a trust assessment method based on crowdsourcing. \cite{li2011multi} presents an algorithm combining WMA-OWA for trust evaluation using five metrics in extensive P2P networks. A broader approach to trust assessment in distributed networks is offered by \cite{4146807}, defining network trust and creating entropy and probability-based models for it. \cite{8616885} investigates three challenges in trust evaluation within VANETs, suggesting an indirect trust evaluation method using reinforcement learning and three trust attributes. Direct and indirect node trust assessment methods, incorporating six trust metrics, are proposed by \cite{mendoza2018distributed}. \cite{chen2011trm} delivers a trust model for IoT systems, applying fuzzy theory and exploring the concepts of trust and reputation within IoT contexts. Both \cite{bao2012dynamic} and \cite{6263792} introduce a dynamic trust management protocol to address malicious behaviours in IoT systems, distinguishing between malicious and non-cooperative nodes and setting stringent criteria. Enhancements to these protocols, including storage management strategies for scalability, are made in \cite{6513398}. \cite{fernandez2017modelling} proposes a sophisticated, layered architectural design aimed at crafting a trust management system for IoT. \cite{8941247} describes a trust and privacy-preserving recommendation scheme for vehicle platoons, involving trust valuation for selecting the lead vehicle.  \cite{9739996} introduces a framework for assessing IoT service trustworthiness, adapting to consumer usage patterns and employing a novel algorithm to detect service-specific trust indicators. 

The major limitation of model-driven methods lies in their specific applicability to certain IoT services or groups, hindered by the diverse characteristics of different IoT services. These services often exhibit unique, context-specific features that vary significantly, challenging the effectiveness of a single model across all scenarios. These model-driven approaches necessitate manual analysis of trust characteristics by domain experts for each distinct IoT service, thereby limiting their flexibility and struggling with accuracy and efficiency in diverse IoT environments. 

\subsection{Centralized Data-Driven IoT Service Trust Prediction}
In the field of centralized IoT service trust prediction, a trust management framework using Support Vector Machine (SVM) was proposed in \cite{lopez2015towards}. It focuses on trust management communication. \cite{7011235} considers trust assessment as a classification problem using non-probabilistic binary SVM classifiers, with an emphasis on social network trust evaluation. EMLTrust \cite{akbani2012emltrust} is a reputation system for Mobile Ad-hoc Networks (MANETs) based on digital signatures, utilizing SVM. Additionally, \cite{8246999} and \cite{cheng2021general} leverage machine learning for device assessment in IoT networks and single-agent trust quantification in multi-agent systems, respectively. 

Centralized methods have significant limitations in IoT service trust prediction, including inadequate high-dimensional information processing, lack of direct communication between nodes in MEC environments, and over-reliance on core networks. These limitations lead to suboptimal performance in large-scale and dynamic settings \cite{9583862}.

\subsection{Distributed Data-Driven IoT Service Trust Prediction}
Several studies focus on developing advanced methodologies for data-driven IoT service trust prediction. \cite{wang2019mtes} proposes a probabilistic graphical model for sensor trustworthiness prediction in IoT and suggests an algorithm to optimize energy use in sensor networks. \cite{zolfaghar2011evolution} devises a social network trust framework using an MLP with echo propagation. Additionally, \cite{8818406} defines a distributed trust prediction model for MEC environments as a Network Lasso problem, using Alternate Direction Method of Multipliers (ADMM) 
for SVM in model training and credibility prediction. This is further refined by \cite{9583862} with the Stochastic ADMM (S-ADMM) method to improve its scalability. Furthermore, \cite{bahutair2021multi} introduces a neural network framework for evaluating service provider trust in crowdsourced IoT. Meanwhile, a deep federated learning method is proposed for enhancing IoT security to identify and manage malicious nodes effectively in \cite{10147216}. The study in \cite{10248268} proposes to employ FedEM for predicting the trustworthiness of IoT services in MEC environments. It addresses the challenge of statistical heterogeneity in trust modeling.

Current IoT trust prediction solutions do not fully address the issue of statistical heterogeneity in distributed MEC environments. For example, the pioneering framework introduced by \cite{9583862} has several areas needing enhancement. Specifically, our experiments reveal that the framework's prediction accuracy is suboptimal. In addition, it necessitates numerous communication rounds and extended time for convergence. The study conducted in \cite{10147216} focuses on dealing with the heterogeneity of data in its analysis. It lacks a detailed exploration of the heterogeneity of mixture distribution. Additionally, there is an absence of discussion regarding scenarios where IoT environments are constrained by limited resources. The statistical heterogeneity arising from mixture distributions has been effectively addressed in \cite{10248268}. However, it does not fully 
mitigate the challenge for
efficient training of trust models in the MEC environment. This is crucial for rapid service selection decisions and service delivery in IoT systems.

\section{Preliminary}
\subsection{Federated Expectation-Maximization} 
The Federated Expectation-Maximization (FedEM) is a unique method within federated multi-task learning. It leverages multi-task learning to understand correlations across different clients, enlarging the node sample size and boosting performance. FedEM assumes each local data distribution of a client is a mix of foundational distributions. It tackles statistical heterogeneity in federated learning by learning shared component models and personalized mixture weights. This approach adapts to data diversity, maintaining high accuracy and fairness while addressing distributional differences \cite{marfoq2021federated}. 

FedEM is based on two main premises:
\textbf{Premise 1}: Each local data distribution $\mathcal{D}_t$ is a blend of $M$ latent distributions $\tilde{\mathcal{D}}_m, 1 \leq m \leq M$, illustrated by the following equation:
\begin{equation}
z_t \sim \mathcal{M}\left(\pi_t^*\right), \quad\left(\left(\mathbf{x}_t, y_t\right) \mid z_t=m\right) \sim \tilde{\mathcal{D}}_m, \quad \forall t \in \mathcal{T}
\end{equation}

\textbf{Premise 2}: For all $m \in \left[M\right]$, $p_m\left(\mathbf{x},y\right) = p_m\left(\mathbf{x}\right)$.

The goal of FedEM is to find the optimal parameters of the components $\Theta^* = \left(\theta_m^*\right)_{1\leq m \leq M}$ and mixture weights $\Pi^* = \left(\pi_t^*\right)_{1 \leq t \leq T}$ by minimizing the negative log-likelihood $f\left(\Theta, \Pi\right)$. This complex problem is typically tackled using the Expectation-Maximization (EM) algorithm, which consists of two steps:

- E-step: The algorithm computes the probability of each data point belonging to each component by updating the latent variable distribution $q_t$ for every data point, as follows:
\begin{equation}\label{update q}
\begin{aligned}
& \text { E-step: } \\ & q_t^{k+1}\left(z_t^{(i)}=m\right) \propto \pi_{t m}^k \cdot \exp \left(-l\left(h_{\theta_m^k}\left(\mathbf{x}_t^{(i)}\right), y_t^{(i)}\right)\right), \\&t \in[T], m \in[M], i \in\left[n_t\right] 
\end{aligned}
\end{equation}

- M-step: The parameters $\{\Theta,\Pi\}$ are updated by maximizing the expected log-likelihood. This includes identifying the optimal component parameters and mixture weights using standard optimization techniques. 
\begin{equation}\label{update pi}
\begin{aligned}
& \text { M-step: } \\ & \pi_{t m}^{k+1}=\frac{\sum_{i=1}^{n_t} q_t^{k+1}\left(z_t^{(i)}=m\right)}{n_t}, \quad t \in[T], \quad m \in[M] \\
\end{aligned}
\end{equation}
\begin{equation}\label{update theta}
\begin{aligned}
& \theta_m^{k+1} \in \underset{\theta \in \mathbb{R}^d}{\arg \min } \sum_{t=1}^T \sum_{i=1}^{n_t} q_t^{k+1}\left(z_t^{(i)}=m\right) l\left(h_\theta\left(\mathbf{x}_t^{(i)}\right), y_t^{(i)}\right), \\&m \in[M]
\end{aligned}
\end{equation}

With these steps, the algorithm iteratively refines the estimated parameters, driving the model towards the optimal solution.

The objective of federated optimization can be interpreted in FedEM as the minimization of the global objective function, namely the negative log-likelihood function $f(\Theta, \Pi)$. This function consists of a weighted average of the local loss functions of all MEC environments.
The global optimization problem can be written as follows:
\begin{equation}\label{objective func}
\begin{aligned}
& \min_{\Theta, \Pi} , f(\Theta, \Pi) \
& , \Theta \in \mathbb{R}^{M \times d}, \Pi \in \Delta^{T \times M}, \forall t \in \mathcal{T}
\end{aligned}
\end{equation}
where $f(\Theta, \Pi)$ is the global objective function, which is the weighted average of the local loss functions of all MEC environments, which can be expressed as:
\begin{equation}\label{local loss func}
f(\Theta, \Pi) = -\frac{1}{n} \sum_{t=1}^T n_t \log p(s_t | \Theta, \pi_t)
\end{equation}
where $f(\Theta, \Pi)$ is the global negative log-likelihood across all MEC environments. $n$ and $n_t$ are the total and per-environment data points respectively. $s_t$ denotes data from the $t$-th environment and $\pi_t$ its mixing weight. $p(s_t | \Theta, \pi_t)$ represents the probability of data $s_t$ given the global parameters $\Theta$ and $\pi_t$.

\subsection{Deep Q Network}
Deep Q Network (DQN) is a Q-Learning variant that utilizes deep learning for function approximation. The key to DQN lies in its ability to make intelligent decisions. It autonomously learns and identifies the optimal strategies in complex environments, enabling effective decision-making across various application scenarios \cite{mnih2013playing}. In the context of federated learning, DQN can evaluate and select clients that bring higher efficiency to model training. This selection is based on the relationship between model weights and a reward function focused on increasing validation accuracy and reducing communication rounds. This capability allows DQN to effectively choose clients that contribute most significantly to the overall learning process. In this regard, it is capable of optimizing training time and resource utilization \cite{9155494}. 

The Q-Learning algorithm can be represented as:

\begin{equation}
Q^*(s, a)=\mathbb{E}_{s^{\prime} \sim \mathcal{E}}\left[r+\gamma \max _{a^{\prime}} Q^*\left(s^{\prime}, a^{\prime}\right) \mid s, a\right]
\end{equation}

Here, $(s, a)$ is the current state-action pair, $r$ represents the immediate reward, $(s^{\prime}, a^{\prime})$ are the next state and action, and $\gamma$ is the discount factor.

However, the complexity of learning the optimal action-value function, $Q^*(s, a)$, increases with high-dimensional spaces or continuous states/actions. DQN overcomes this by using a deep neural network, $Q(s, a; \theta)$, approximating the $Q^*(s, a)$ function. This network's parameters, $\theta$, are learned by minimizing a series of loss functions, $L_i(\theta_i)$:

\begin{equation}\label{dqn_loss}
L_i\left(\theta_i\right)=\mathbb{E}_{s, a \sim \rho(\cdot)}\left[\left(y_i-Q\left(s, a ; \theta_i\right)\right)^2\right]
\end{equation}

The target value, $y_i$, is defined as:

\begin{equation}
y_i = r + \gamma \max _{a^{\prime}} Q\left(s^{\prime}, a^{\prime}; \theta_{i-1}\right)
\end{equation}

DQN employs Experience Replay to disrupt continuous sample correlation, thus making the training more stable and efficient. The agent's experiences are stored in a replay buffer, $\mathcal{M} = \{e_1, ..., e_t\}$, and minibatches of these experiences are sampled during the learning process to update the network parameters.

\section{Problem formulation}
In this section, we present the mathematical representation of the problem related to  IoT service trust prediction within statically heterogeneous MEC environments. This formulation enables us to systematically address the issue and develop efficient solutions.

Let us contemplate a set of arbitrary trust-related attributes. These attributes could be any parameters or characteristics that affect the reliability of an IoT service. We organise these trust attributes into a vectorized format, denoted as $x_i$ in $\mathbb{R}^d$, where $d$ signifies the dimension of the $x_i$ vector. In simpler terms, $d$ represents the number of distinct trust attributes we are considering.

Each trust attribute possesses a certain level of significance or influence on the overall trustworthiness value. This is expressed through a coefficient vector $w$ in $\mathbb{R}^d$. Each element of $w$ corresponds to the weight or impact of the respective trust attribute in the $x_i$ vector.

The function $tr$ is defined as a mapping function that combines each trust attribute with its corresponding weight coefficient. The result of this function provides us with the overall trust value $y$.

Hence, the mathematical representation of the trust model for any arbitrary instance can be expressed as follows:
\begin{equation}
\hat{y_i}=\operatorname{tr}\left(x_i ; w\right) \quad \text { where } \quad \operatorname{tr}: \mathbb{R}^d \times \mathbb{R} \Rightarrow \mathbb{R}
\end{equation}
where $\hat{y_i}$ represents the predicted trust value for the $i$-th instance. The function $\operatorname{tr}$ denotes the mapping operation that takes a vector from $\mathbb{R}^d$ and a scalar from $\mathbb{R}$ as inputs and produces a scalar in $\mathbb{R}$ as the resulting overall trust value calculation.

Specifically, the trust prediction model under consideration is designed for a topology consisting of multiple MEC environments. Let $\mathcal{T}=\left\{t_1, t_2, \ldots, t_m\right\}$ denote a collection of MEC topologies, with $m$ MEC environments in total. Each of these MEC environments possesses its own local dataset, denoted as $D=\left\{d_1, d_2, \ldots, d_m\right\}$, containing data from all $m$ MEC environments. Due to the varying characteristics of these environments, the local data distribution, denoted as $\left\{\mathcal{D}_t\right\}_{t \in \mathcal{T}}$, is typically non-uniform. As a result, each data distribution $\mathcal{D}_t$ is used to train separate model weights, represented as $w_{\mathcal{D}_t} \in \mathcal{W}$. The ultimate objective is to find the optimal set of $w$ values, which usually involves minimizing a loss function. This can be mathematically expressed in the objective function as follows:

\begin{equation} \label{objective func}
\mathop{minimize}\limits_{w \in \mathcal{W}}\mathcal{L}_{\mathcal{D}_t}(w)=\frac{1}{\mathcal{T}}\sum_{t=1}^{\mathcal{T}} \frac{n_t}{n} \mathbb{E}_{\mathcal{D}_t} \mathcal{L}_{t_{\mathcal{D}_t}}(w)
\end{equation}

In this equation, $n_t$ denotes the size of the local dataset in the $t$-th MEC environment. $\mathbb{E}_{\mathcal{D}_t}$ represents the expected value computed from a sample of the dataset $\mathcal{D}_t$ located in MEC environment $t$. Additionally, $n$ denotes a round of sampling from the overall dataset.

A critical aspect of this procedure involves the identification of potential weight relationships among MEC environments based on the disparities in their local data distributions. For every distinct MEC environment within the set $\mathcal{T}$, a local data distribution $\mathcal{D}_t$ is generated from its respective local dataset. The specific local loss function pertaining to the $t$-th MEC environment and its unique data distribution is represented as $\mathcal{L}_{t_{\mathcal{D}_t}}$. For each individual MEC environment, this local loss function, denoted as $\mathcal{L}_t$, can be further elaborated in the following manner:
\begin{equation}
\mathcal{L}_t(w)=\frac{1}{n_t} \sum_{i=1}^{n_t} \ell_i(w)
\end{equation}
This equation represents the local loss function, where $\ell_i$ stands for the individual loss function associated with each data sample within the specific local MEC environment. In this context, $n_t$ corresponds to the size of the local dataset for the $t$-th MEC environment, and $w$ represents the weight coefficients.

The local model within each MEC environment is updated using the following equation:
\begin{equation}
w_{r+1}^t=w_r^t- \eta \nabla \mathcal{L}_t(w)
\end{equation}
where $\eta$ denotes the learning rate, which is a parameter that governs the magnitude of adjustments made to the weights of the model in response to the loss gradient. The symbol $\nabla$ represents the gradient operation, indicating the direction of the steepest ascent of the function. The notation $w_{r+1}^t$ represents the updated weight vector for the $t$-th local model in the subsequent round of iteration.

Once the local models in the MEC environments have been updated, the trained model weights are transmitted to the central cloud. Subsequently, the central cloud aggregates these weights while taking into account the distinct MEC environments. The aggregation operation can be represented as follows:
\begin{equation} \label{aggregates}
w_{r+1}=\sum_{t=1}^{\mathcal{T}} \frac{n_t}{n} w_{r+1}^t
\end{equation}
This equation demonstrates how the central cloud computes a weighted sum of the model weights from the various MEC environments for the upcoming round. The weights used for aggregation are proportional to the size of the dataset in each respective MEC environment. The resulting aggregated weight vector, denoted as $w_{r+1}$, is then employed in the subsequent round of the model training process across all MEC environments.

\section{Proposed Solution}
This section provides a comprehensive overview of the proposed solution and the theoretical foundations on which it was developed. Section 5.1 offers a complete and smoothly articulated explanation of the proposed solution to ensure easy understanding. Section 5.2
provides a comprehensive and detailed introduction to our proposed solution.

\subsection{Overview of FEDQ-Trust}
Our proposed solution, FEDQ-Trust, leverages a federated multi-task learning algorithm for trust prediction in MEC-based IoT systems. We integrate this algorithm with a DQN model to enhance efficiency and performance. The core concept of FEDQ-Trust is to address trust prediction as a federated optimization problem aimed at minimizing the global negative log-likelihood (Equation \ref{objective func}). 

FEDQ-Trust operates on the premise that the trust prediction model of each MEC environment should be trained on its local data distribution to provide accurate predictions while preserving the privacy of local data. This recognition arises from the fact that IoT service trust data in different MEC environments may exhibit distinct data distributions. The integration of DQN into our solution facilitates efficient environment selection during training, optimizing the overall training process and accelerating convergence.

Our proposed algorithm is implemented within a distributed architecture involving decentralized MEC environments and a central cloud. In this architecture, the local trust prediction parameters of the model within each MEC environment are transmitted to the central cloud through a bidirectional communication link. This enables federated learning, where local model parameters are aggregated in the central cloud to create a global trust prediction model capable of adapting to variations among MEC environments.

Each MEC environment performs two critical tasks: EM updates for component weights optimization and local model training on locally aggregated data distributions. This distributed federated learning strategy consolidates data from diverse MEC environments in the central cloud, resulting in more accurate models. Our data-driven approach enhances trust prediction accuracy and can be applied across various distributed IoT systems in MEC environments. In summary, our algorithm achieves distributed trust prediction by enabling two-way communication between the global cloud and each MEC environment, leading to more precise and reliable trust predictions.

The algorithm follows a workflow consisting of six key steps, as depicted in Figure \ref{fig:visualize}, outlining the federated optimization process:
\begin{figure*}[htbp]
    \centering
    \includegraphics[width=17cm]{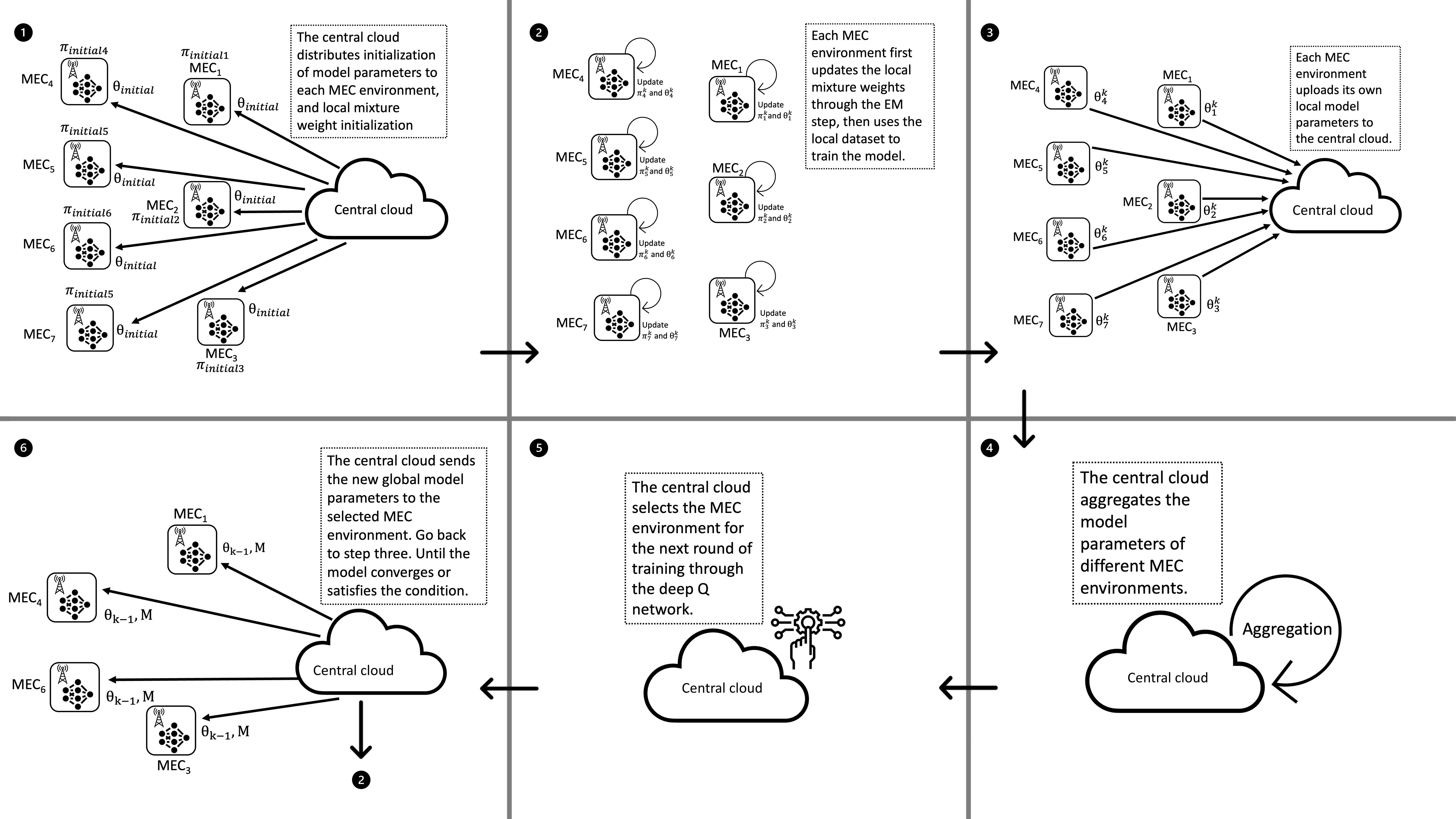}
    \caption{A visualization of the information flow associated with the federated optimization for MEC-based IoT systems.}
    \label{fig:visualize}
\end{figure*}

\begin{itemize}
\item[1)] 
Initialization: The initial step broadcasts the weights of the global trust prediction model from the central cloud to each MEC environment and the initialization of the local mixture weights by each MEC environment. Furthermore, the local mixture weights initialization accommodates data distribution of each respective MEC environment, influencing the upcoming model training and parameter updates.

\item[2)] 
EM Updating and Local Training: In the second step, we require each MEC environment to perform an update on their locally optimized component weights, a process carried out through the Expectation-Maximization (EM) step. Concurrently, every MEC environment will employ its locally optimized component weights and local datasets for model training, followed by the transmission of these trained model weights back to the central cloud for the next round of aggregation. It is during this step that each MEC environment implements updates on their local mixture weights through the EM step (See Equations 8-10), thereby tailoring it to better fit the local dataset.

\item[3)] 
Local Model Uploading: The third step involves each MEC environment uploading its locally trained model weights to the central cloud for model aggregation. 

\item[4)] 
DQN-Based Environment Selection: The fourth step introduces a DQN model to select the optimal MEC environment. The DQN model is responsible for deciding which MEC environments will receive the updated global model weights in the following step, improving the efficiency of the federated learning process.

\item[5)] 
Aggregation: In the fifth step, after the central cloud receives the model weights of the local MEC environment, the central cloud aggregates them according to the model weights of each MEC environment, thus completing a round of communication. In this step, the central cloud will receive model weights from multiple MEC environments and aggregate them based on the weights of each MEC environment. This helps synthesize information from multiple environments and derive parameters for a global model (see Equation \ref{aggregates}).

\item[6)] 
Local Model Updating: For the final step, only the MEC environments selected by the DQN model in step four receive the aggregated global weights. These environments simultaneously update their local mixture weights and local model weights before sending this updated data back to the central cloud. This marks the beginning of the next iteration, which will continue until optimal model parameters are obtained, or a predetermined stopping condition is reached.
\end{itemize}

\begin{algorithm}[h]
\SetKwFunction{FT}{EMStepWithAggregate}
\SetKwInOut{Input}{Input}
\SetKwInOut{Output}{Output}
    \caption{Procedure of FEDQ-Trust}
    \label{alg:algorithm1}
    \Input{Set of MEC environments $\mathcal{T}$; Set of local MEC environment data $D_{1:\mathcal{T}}$; Number of mixture distributions $M$; Number of communication rounds $K$; Target Accuracy $A$} 
    \Output{Model parameters $\theta_m^K$, $m \in \left[ M \right]$}
    \begin{algorithmic}[1]
        \STATE \emph{Central cloud initialize replay memory $\mathcal{M}$, initial Q-values and state-action pair}\;
        \STATE \emph{Central cloud initialize $\theta_{0}$ for $1 \leq m \leq M$ to the $\mathcal{T}$ all MEC environments}\;
        \FORALL{$t = 1$ to $\mathcal{T}$ all MEC environments}
            \STATE \emph{initialize $\pi_t$}\;
        \ENDFOR
        \STATE \FT{}\;
        \FORALL{$m = 1$ to $M$}
            \STATE Perform PCA on $\theta_{m}$ to reduce dimensionality
            \STATE $s_{init}^{m} \leftarrow$ reduced-dimension $\theta_{m}$
        \ENDFOR        
        \FORALL {$k = 1$ to $K$}                
            \STATE {\emph{Calculate Q-values and select a action $a_t$ based on the softmax of Q-values' probabilities}}\;
            \STATE {\emph{Central cloud broadcasts $\theta_{k-1, m}$ for $1 \leq m \leq M$ to the selected MEC environments by DQN}}\;
            \STATE {\emph{\FT{}}}\;
            \STATE Perform PCA on $\theta_{k, m}$ for $1 \leq m \leq M$ to reduce dimensionality
            \STATE $s_{k+1}^{m} \leftarrow$ reduced-dimension $\theta_{k, m}$ for $1 \leq m \leq M$
            \STATE {\emph{$r_k \leftarrow \Xi^{A_k - A} - 1$}}\;
            \STATE {\emph{Store transition $(s_k, a, r_k, s_{k+1})$ in $\mathcal{M}$}}\;
            \STATE {\emph{Sample random minibatch of transitions $(s_j, a, r_j, s_{j+1})$ from $\mathcal{M}$}}\;
            \STATE {Set $y_j \leftarrow \begin{cases} r_j & \text{for } A_t \geq A  \\  r_j+\gamma \max _{a^{\prime}} Q\left(s_{j+1}, a^{\prime}\right) & \text{for } A_t < A \end{cases}$}
            \STATE {\emph{Perform a gradient descent step on $(y_j - Q(s,a))^2$}}\;
        \ENDFOR
    \end{algorithmic}  
    \BlankLine
    \SetKwProg{Func}{Function}{:}{}
    \Func{\FT{}}{}
    \begin{algorithmic}[1]
        \FORALL{$t = 1$ to selected $T$ partial MEC environments}
            \FORALL{$m = 1$ to $M$}
                \STATE //  E-step:
                \STATE update $q_t^k\left(z_t^{(i)}=m\right)$ as in $(\ref{update q})$, $\forall i \in\left\{1, \ldots, n_t\right\}$ 
                \STATE //  M-step:
                \STATE update $\pi_{t m}^k$ as in $(\ref{update pi})$
                \STATE $\theta_{t m} \leftarrow LocalSolver (m, \theta_{t m}, q_t^k, D_t)$
            \ENDFOR
            \STATE \emph{MEC environment $t$ sends $\theta_{tm}$ for $1 \leq m \leq M$ to the central cloud}
        \ENDFOR
        \FORALL{$m = 1$ to $M$}
            \STATE $\theta_{m} \leftarrow \sum_{t=1}^T \frac{n_t}{n}\cdot\theta_{t,m}$
        \ENDFOR
    \end{algorithmic}
\end{algorithm}
\subsection{Technical Details of FEDQ-Trust}
Our study addresses the issue of heterogeneity in MEC-based IoT systems. This heterogeneity results from different data distributions and non-identically distributed (non-IID) data. We use the FedEM algorithm to manage this challenge \cite{marfoq2021federated}. The FedEM algorithm is known for dealing effectively with non-IID and mixed data in federated learning. It performs Expectation-Maximization (EM) steps in each MEC environment. These steps help in federated optimization. They also improve prediction abilities in various environments. Section 3.1 provides more details about the FedEM algorithm.

Our framework uses a multi-layer perceptron (MLP) as its basic learning model. The MLP has an input layer, a hidden layer with many perceptrons using ReLU functions, and an output layer. In the output layer, a sigmoid function activates one perceptron. Each MEC environment trains its local MLP as a binary classifier. The performance of these classifiers differs because of the various data distributions they face. However, we use federated optimization to combine them into a global model for trust prediction.

Our methodology further refines the efficiency of convergence of FedEM by integrating a reinforcement learning approach powered by a DQN agent. This DQN agent intelligently selects a specific subset of MEC environments in each iteration of federated learning, optimizing the convergence speed while maintaining model accuracy. The parameters of the global model constitute the state of DQN, which we simplify using Principal Component Analysis (PCA) due to the inherent high dimensionality of parameters in deep learning applications \cite{9155494}. Within the FedEM framework, individual models are tailored to each distribution, followed by PCA application to condense the model parameters into a manageable state space. The DQN is rewarded based on the accuracy of the current iteration and a predefined accuracy. The technical details of DQN are explained in Section 3.2.

Our DQN-enhanced FedEM strategy is outlined in Algorithm \ref{alg:algorithm1}. First, the central cloud server prepares the replay memory and sets the initial model parameters. Then, each MEC environment starts its own setup, as shown in lines 1-5 of the algorithm. Before the main training begins, we run an initial FedEM training cycle. This first cycle sets the starting state of the DQN model, as mentioned in line 6. After this, we apply PCA to the parameters from both global and local models that finished the initial FedEM training. This process simplifies the state space, as described in lines 7-10 of the algorithm.

The global model's weights help choose an action from the MEC environment pool once DQN training begins. This selection process is in line 12 of the algorithm. After this choice, we update the model parameters and send them to the selected MEC environments. This starts a new FedEM training cycle, shown in line 13. Then, these local environments train with their data. The central cloud combines these updated local models, as line 14 indicates. Next, we apply PCA to the new global model. This compresses its parameters, forming the next state for the DQN, as lines 15-16 describe. We calculate the reward by comparing the target and current accuracy. In our tests, we set the accuracy threshold $\Xi$ at 64, as line 17 states.

The tuple that includes the current state, the action taken, the resulting reward, and the next state is saved in the replay memory, as line 18 of the algorithm shows. The algorithm continues by choosing random transition samples from the replay memory for the DQN network's training phase. This is line 19 of the algorithm. Line 20 deals with how to calculate the target value $y_j$. If the observed accuracy $A_t$ meets or exceeds the set accuracy goal $A$, then $y_j$ equals the current reward $r_j$. If $A_t$ is below $A$, $y_j$ is the sum of $r_j$ and the decay factor $\gamma$ multiplied by the highest forecasted Q value for all possible actions $a'$ in the next state $s_{j+1}$. Line 21 carries out an optimization step. It uses gradient descent to reduce the difference between the expected Q value $y_j$ and the Q value $Q(s,a)$ predicted by the network. This matches the loss function in Equation (\ref{dqn_loss}). The procedure in line 11 repeats until the algorithm completes the set number of training cycles. It ends with the final model parameters for prediction.

Our algorithm uses a process called "EMStepWithAggregate" during federated optimization. It works at the same time across the chosen $T$ subsets of MEC environments. This process carries out Expectation (E) and Maximization (M) steps for each part of the mixture distribution. In the E-step, the algorithm updates the posterior probabilities for hidden aspects of each data sample. This is detailed in line 4 of the EM algorithm and follows the updating rule in Equation (\ref{update q}). In the M-step, it changes the weights and parameters of the mixture components. This is based on lines 6 to 7 by following the formulas in Equations (\ref{update pi}) and (\ref{update theta}). Each MEC environment sends its new parameters to the central cloud after updating. This is shown in line 9 of the EM algorithm. The central cloud then combines these parameters. It does a proportional aggregation as described in lines 11 to 13 of the EM algorithm. This combination creates the global model parameters.

\section{Evaluation}

Our experimental evaluations were conducted on a computer equipped with an Apple Silicon M1 Pro 10 Cores processor and 32GB of RAM. All comparison models were implemented using the Python programming language. To support the implementation of federated learning models, including FedTrust and FedEM, as well as the DQN component in our experiments, we utilized the PyTorch (version 2.0.1) library \cite{9155494} \cite{10147216} \cite{marfoq2021federated}. Additionally, the CVXPY (version 1.3.2) library was utilized to implement other MEC-based IoT service trust prediction models, namely S-ADMM and Local SVM \cite{9583862}.
The hyperparameter values for all models were meticulously tuned to achieve their optimal performance.
We implemented all the models in Python and utilize the aforementioned libraries. This allows us to conduct our experiments in a standardized and reproducible manner, ensuring the accuracy and reliability of our results. The source code of our proposed solution can be accessed at $\footnote{https://github.com/SHVleV9CYWkK/FEDQ}$,
along with the simulations linked to these experiments.

\subsection{Experiments}
A series of experiments were conducted to thoroughly assess the effectiveness of the proposed method in predicting the trustworthiness of IoT services in MEC-based IoT systems. 
The conducted experiments are elaborated as follows.

\begin{itemize}
\item
\textit{Accuracy} stands as a key performance indicator for any machine learning approach. This metric is particularly essential, as it quantifies the percentage of samples within MEC scenarios that are accurately forecasted by the algorithm. This metric serves as a benchmark to evaluate the comparative effectiveness of various models in predicting the trustworthiness of IoT services. 


\item
\textit{Training Time Comparison} provides insight into the computational speed of our method relative to others. This metric represents the complete duration required for a model to achieve convergence in its training phase. By evaluating the elapsed time across different models, we can determine which one operates most efficiently. This is especially vital in extensive MEC environments, where the speed of computation plays a major role in overall performance.


\item 
\textit{Convergence Iterations Comparison} serves as a crucial metric. It measures data exchange during model training. The metric counts interactions between the central server and local devices. These interactions continue until the model converges. The measure identifies models with optimal performance and minimal communication rounds. It also shows how much the DQN component reduces computational complexity. This factor is important in MEC environments with limited communication resources.

\item 
\textit{Comparison of DQN Component and Random Selection} focuses on two key aspects. First, it looks at how intelligent MEC environment selection methods with DQN components differ from traditional random selection methods in resource selection efficiency. Second, it aims to validate the effectiveness of the DQN component. The contrast with random selection highlights the advantages of the DQN component.
\end{itemize}

\subsection{Compared Models}

The following models are selected as the baseline for comparison. 

 \textbf{FedTrust} \cite{10147216} is a federated learning-based IoT service trust prediction approach. It excels in identifying harmful nodes in MEC-like environments using a trust dataset enriched with knowledge, experience, and reputation. 
 
\textbf{S-ADMM} \cite{9583862} is another state-of-the-art data-driven IoT service trust prediction approach. It can transform the distributed trust prediction problem into a network lasso problem.

\textbf{Local SVM} \cite{8818406} conducts IoT service trust prediction independently in each MEC environment.

\subsection{Datasets}
The following real-world IoT security datasets are leveraged for our evaluation.

UNSW-NB15 Dataset$\footnote{https://research.unsw.edu.au/projects/unsw-nb15-dataset}$: This dataset comprises an extensive range of network traffic data. It includes representations of both standard network operations and artificial attack activity, which were synthesized in a controlled lab environment. The dataset encompasses approximately 2 million entries, each detailed with 49 distinct features. Every entry aligns with an outcome, categorized as either a normal transaction or one among nine varied attack types. In our experimental context, these nine attack types were classified as malicious, while all standard transactions were labelled as non-malicious.

N-BaIoT Dataset$\footnote{https://archive.ics.uci.edu/ml/datasets/detection\_of\_IoT\_botnet\_\newline attacks\_N\_BaIoT}$: This dataset is an aggregation of network traffic data, sourced from compromised real commercial IoT devices. It features roughly 7 million entries, with each entry detailed by a comprehensive set of 115 features. In this dataset, attack types are broadly divided into two main categories, each encompassing various specific types of attacks. All instances of these attacks are labelled as malicious, whereas the rest of the data, representing normal operations, are identified as benign.

We allocated 80\% of the data for training and 20\% for testing to maintain a consistent and unbiased evaluation of the models. We divided the datasets into 100 local datasets of MEC environments. The sizes of these local datasets differ considerably, showcasing a range of data distributions that mirror the diverse nature typically seen in real-world MEC systems. The training dataset sizes range from 5 to 613,791 entries, while the test dataset sizes span from 19 to 215,016 entries. In federated learning with artificial neural network (ANN) models, a trust value of 0 means benign instance, and 1 means malicious instance. This differs from other models that use 1 for benign instances and -1 for malicious instances.

We systematically segmented the dataset and generated a mixture distribution before model training. This process was crucial to ensure uniformity across all models. This approach guarantees that our experimental dataset aligns with the mixture distribution hypothesis. Initially, the dataset is segmented into two distinct clusters. Within each cluster, the samples are then distributed across 100 MEC environments, guided by a Dirichlet Distribution with an alpha value of 0.4\footnote{
The work in \cite{10248268} adopts an alpha value of 0.3, resulting in no positive examples in a few local MEC environments. Consequently, the experimental results are different from those presented in \cite{10248268}. 
}. For each MEC environment, sample selection from each cluster is executed in a manner that aligns with the Dirichlet distribution of the corresponding cluster. Specifically, the Dirichlet distribution is used as a prior for the multinomial distribution, and Bayesian inference updates the posterior distribution with each new sample. The update forms a new probability distribution. This results in a mixture distribution. It shows the diverse characteristics of the data in MEC environments \cite{marfoq2021federated}.

\subsection{Key Performance Indicators}
We assess the efficacy of our FEDQ-Trust model. We compare it with the baseline models depicted in Section 6.2 with the following Key Performance Indicators (KPIs).

\textit{Accuracy} stands as a pivotal measure in the realm of machine learning, holding particular significance in our study. This metric quantifies the ratio of correctly predicted instances in the MEC environments relative to the model's predictions. Accuracy is the number of correctly predicted instances divided by the total number of samples.

\textit{True Positive Rate (TPR)}, also known as sensitivity, is very important in our analysis. This is because our datasets have an imbalance of positive and negative samples. TPR evaluates the proportion of actual positive instances (true positives) accurately identified by the model from all actual positive instances present. This rate is determined by dividing the number of true positive outcomes by the total of true positives plus false negatives. 

\textit{False Positive Rate (FPR)}, also known as the probability of false alarm, assumes a significant role in our study. This importance comes from the uneven distribution of positive and negative samples in our datasets. FPR measures how often the model wrongly identifies negative instances as positive. It compares these cases to all actual negative instances. This metric is calculated by dividing the number of false positives by the sum of false positives and true negatives. 

\textit{Communication Rounds} serves as a key metric to gauge the data exchange load involved in the training of models. This count reflects the number of interactions required between the central server and local devices for a model to reach convergence. 

\textit{Elapsed Training Time} is a important KPI. It measures the training efficiency of our algorithm. This is important for comparing it with other models. This metric is defined as the total time a model requires to reach convergence throughout the training phase. 


\subsection{Results and Discussion}

\subsubsection{Accuracy}
\begin{figure}[htp]
    \centering
    \includegraphics[width=6cm]{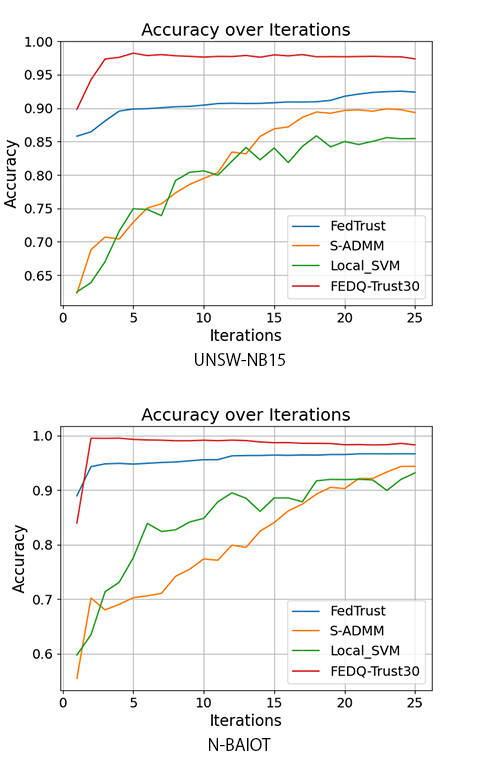}
    \caption{Accuracy of each model on the two datasets
    }
    \label{fig:test-acc}
\end{figure}

The choice of FEDQ-Trust30 for this experiment stems from experimental observations. These observations show FEDQ-Trust30 as the most optimized model in intelligent environments. This comparison and the decision are detailed in Section 6.5.4.
These results, as depicted in Figure \ref{fig:test-acc}, clearly highlight the performance advantages of FEDQ-Trust. The graph illustrates not only the superiority of FEDQ-Trust in terms of accuracy but also its consistency across different datasets. This consistency is crucial, considering the varied nature of the datasets. These results underscore the superior predictive capability of FEDQ-Trust in diverse testing scenarios.

Especially, FEDQ-Trust30 showed remarkable performance on the UNSW-NB15 dataset. It achieved an impressive 97.59\% accuracy in just the 4th round. In contrast, FedTrust and S-ADMM only managed to reach accuracies of 92.5\% and 89.9\%, respectively. 
The performance of the Local SVM model fell short of expectations with an accuracy rate of 85.5

Similarly, on the N-BaIoT dataset, FEDQ-Trust30 continued to showcase its robustness, attaining the highest accuracy of 99.5\%. In the same setting, FedTrust's performance was slightly lower. It achieved an accuracy of 96.6\%. S-ADMM also performed lower with an accuracy of 94.3\%. Meanwhile, Local SVM reached an accuracy of 93.1\%.

The outstanding performance of FEDQ-Trust is attributed to its ability to find mixture weights for different local data distributions, which facilitates efficient processing of diverse and complex data structures and intelligent selection of the most contributive training environments. Its design enables it to quickly adapt and learn from a variety of datasets, thereby achieving high accuracy in early rounds.

\subsubsection{True Positive Rate}
\begin{figure}[htp]
    \centering
    \includegraphics[width=6cm]{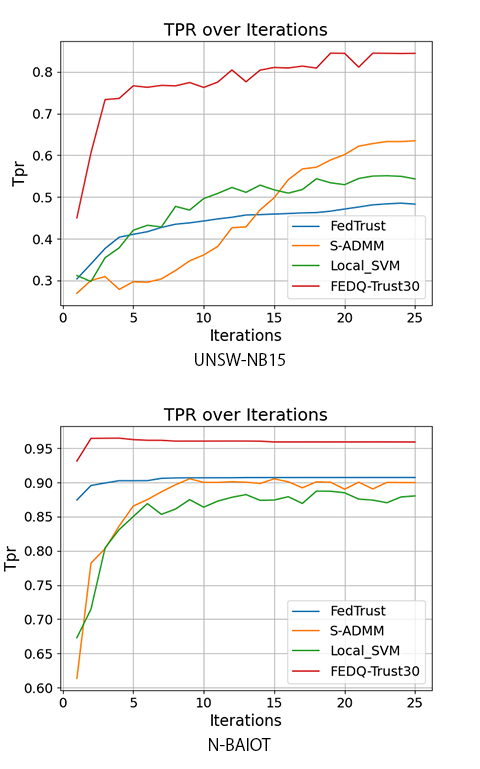}
    \caption{True Positive Rate of each model on the two datasets
    }
    \label{fig:test-tpr}
\end{figure}

The TPR data is shown in Figure \ref{fig:test-tpr}. It highlights the ability of FEDQ-Trust. FEDQ-Trust consistently identifies true positives effectively. Within the UNSW-NB15 dataset, FEDQ-Trust30 exhibits a rapid increase in TPR. It maintains this high TPR throughout the iterations. This indicates a strong and consistent ability to correctly identify true positives. Its performance exceeds that of both FedTrust and Local SVM demonstrating a slower growth and lower overall TPR. S-ADMM displays a more gradual increase, suggesting a steadier but less pronounced enhancement in identifying true threats compared to FEDQ-Trust30.

FEDQ-Trust30 starts with a strong performance on the N-BaIoT dataset. It quickly stabilizes at a high TPR, demonstrating its effectiveness in identifying true positives. However, FedTrust exhibits a marginally higher TPR after initial iterations, indicating a comparable detection capability. The Local SVM improves over time. However, it does not reach the TPR levels of FEDQ-Trust30 or FedTrust. This suggests limitations in its detection accuracy. S-ADMM shows gradual improvement but plateaus below the TPR of the FEDQ-Trust30 and FedTrust.

These findings demonstrate the effectiveness of FEDQ-Trust. It accurately identifies true positives. This is a crucial attribute for any security tool in IoT network environments. Such consistent performance in recognizing threats bolsters the reliability of FEDQ-Trust as a significant component of IoT security.

\subsubsection{False Positive Rate}
\begin{figure}[htp]
    \centering
    \includegraphics[width=6cm]{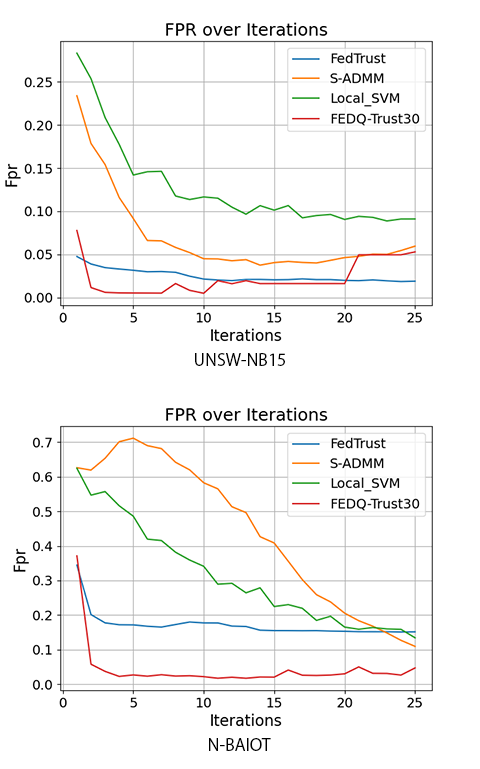}
    \caption{False Positive Rate of each model on the two datasets
    }
    \label{fig:test-fpr}
\end{figure}

FEDQ-Trust30's performance in terms of FPR in Figure \ref{fig:test-fpr} is commendable. FEDQ-Trust is compared to other models. It shows a lower FPR. This indicates fewer false positives. This is different from models like FedTrust and Local SVM. They exhibit higher FPRs. This suggests they are more prone to incorrectly flag benign activities as threats. The FPR of S-ADMM is also higher than FEDQ-Trust30, but it shows a steady decline, indicating improving accuracy over iterations. Overall, FEDQ-Trust30 demonstrates a balance between maintaining low false positives and achieving high true positives.

FEDQ-Trust30 again demonstrates a low FPR on the N-BaIoT dataset, indicative of its accuracy in avoiding false alarms. FedTrust initially has a higher FPR that decreases over time. In contrast, FEDQ-Trust30 starts with a low FPR and remains low. This showcases its consistent precision. S-ADMM's FPR starts high. However, it shows a significant decrease over time. This reflects an improvement in identifying true negatives as iterations progress. Local SVM starts with the highest FPR. It does experience a drop over time. However, it ends with a higher FPR than FEDQ-Trust30. This highlights its challenges in accurately classifying negative instances.

The graphs depicting FPR across iterations for both datasets clearly establish FEDQ-Trust as the leading model with respect to minimizing false positives. This not only affirms the efficiency of FEDQ-Trust in discerning true negatives from false positives but also highlights its potential in practical applications where the cost of false alarms can be substantial. 

\subsubsection{Performance of DQN Components}
\begin{figure}[htp]
    \centering
    \includegraphics[width=6cm]{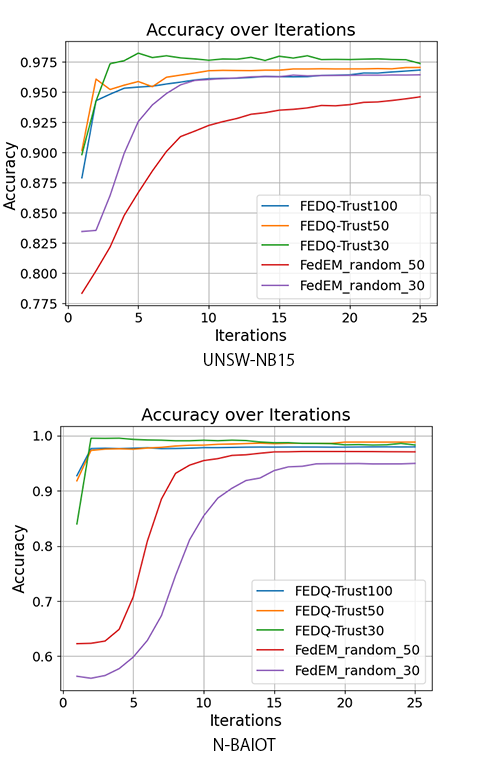}
    \caption{Accuracy of selection component for MEC environments on two datasets
    }
    \label{fig:dqn-acc}
\end{figure}

Figure \ref{fig:dqn-acc} paints a clear picture of the effectiveness of the DQN components in our FedEM models. 
FEDQ-Trust100 is the FedEM method without DQN-based MEC environment selection. FEDQ-Trust50 denotes a model where 50 MEC environments are selected each round. FEDQ-Trust30 denotes a model where 30 MEC environments are selected each round. In both, the selection is done intelligently through a DQN component for the training process. Meanwhile, FedEM\_random\_50 and FedEM\_random\_30 are methods with a random approach. They randomly select 50 and 30 MEC environments per round, respectively.

FEDQ-Trust30 outperforms other configurations, suggesting that selecting 30 well-informed MEC environments is optimal for model training. FEDQ-Trust50, performing closely to the best model, shows that choosing 50 MEC environments captures sufficient diversity for effective learning, though it might include less informative environments, slightly affecting the model's performance. FEDQ-Trust100, ranking third, demonstrates that including more environments offers a wider data range, but can also introduce noise or irrelevant information, potentially hindering the learning of critical patterns.

Both 50 and 30 randomly selected MEC environments show a delayed improvement in accuracy over iterations. This delay in reaching higher accuracy levels suggests something. Random selection may need more iterations to cover the informative diversity of MEC environments. The DQN-based selection achieves this more directly. These models start with lower performance. This could be due to initially including a less optimal mix of environments. Over time, the global model gets exposed to more varied data from random selections. Then, their performance begins to converge with the selectively composed models.

These observations highlight the potential benefits of selectively including MEC environments in the training process. The superior performance of FEDQ-Trust30 shows that a carefully chosen subset of environments provides a robust training dataset. This enables the global model to learn and generalize quickly.

\begin{table*}[h]
    \centering
    \label{tab:table1}
    \caption{Performance comparison among all the models}
    \begin{tabularx}{\textwidth}{XXXXX}
\hline
Dataset                    & Model  & Maximum Accuracy &Elapsed Training Time & Number of Communication Rounds \\ \hline
{UNSW-NB15} & FedTrust & 92.46\%     &325.28s & 23                   \\
                           & S-ADMM & 89.64\% &625.32s &20 \\
                           & LocalSVM & 85.58\% &720.66s &23 \\
                           & FEDQ-Trust100  & 96.57\%     &621.17s & 21                    \\   
                           & FEDQ-Trust50  & 96.57\%     &116.25s & 9                    \\   
                           & FEDQ-Trust30  & 97.59\%     &18.71s & 4                    \\   \hline
                        
{N-BaIoT}   & FedTrust & 96.45\%     &673.86s & 17                    \\
                           & S-ADMM & 94.35\% &2205.29s& 24 \\  
                           & LocalSVM & 93.17\% &2564.07s &25 \\
                           & FEDQ-Trust100  & 97.64\%  &199.7s   & 2            \\ 
                           & FEDQ-Trust50  & 98.56\%  &452.19s   & 13            \\ 
                           & FEDQ-Trust30  & 99.52\%  &39.52s   & 2            \\  \hline    
\end{tabularx}
\end{table*}

\subsubsection{Convergence}
Our analysis of the convergence properties of various models is detailed in Table 1. It provides a clear demonstration of the FEDQ-Trust method's superior convergence characteristics in terms of accuracy, elapsed training time, and communication rounds.

The FEDQ-Trust30 model achieved the highest accuracy on the UNSW-NB15 dataset. It reached a maximum of 97.59\%. This was done in the shortest elapsed time, only 18.71 seconds. It also required the fewest communication rounds, just four. This is an indication of FEDQ-Trust's efficient learning and communication protocol. FEDQ-Trust50 showed remarkable efficiency. It matched the maximum accuracy of FEDQ-Trust100. It converged in nearly one-fifth of the time. It also required less than half of the communication rounds. Comparatively, FedTrust, S-ADMM, and LocalSVM exhibited lower maximum accuracies of 92.46\%, 89.64\%, and 85.58\% with significantly longer elapsed times and more communication rounds. This suggests that they are less efficient and consume more computing resources and time.

The FEDQ-Trust30 model reached perfect accuracy (99.52\%) for the N-BaIoT dataset. This was done in the least amount of time, only 39.52 seconds. It also used the least number of communication rounds  (2 rounds).
FEDQ-Trust50 achieved a 98.56\% accuracy. This shows a high degree of learning effectiveness. However, it required more time and communication rounds than FEDQ-Trust30. FEDQ-Trust100 attained a high accuracy of 97.64\%. It needed more time and rounds than FEDQ-Trust30 and FEDQ-Trust50. FedTrust, S-ADMM, and LocalSVM showed lower maximum accuracies and required considerably more time and communication rounds to converge.

The pattern across both datasets leads to a conclusion. FEDQ-Trust methods, especially FEDQ-Trust30, provide higher accuracy. They also require less time and fewer communication rounds to converge. This efficiency is crucial in MEC environments, where computational resources and time are of the essence. The FEDQ-Trust method has a clear advantage. It converges swiftly to higher accuracy levels with minimal communication overhead. This showcases its suitability for real-time, high-stakes applications. In these applications, quick and accurate decision-making is vital.

\section{Conclusions and future work}
The study introduces FEDQ-Trust, a new trust prediction method for IoT services, which uses federated learning to tackle the challenges of mixture distributions in distributed MEC environments. The training of trust prediction models in MEC is framed as a federated optimization problem. This involves integrating a FedEM approach to tackle mixture distributions and a DQN-based reinforcement learning technique for the smart selection of MEC environments. The method's practicality, effectiveness, and superiority over existing IoT service trust prediction methods for MEC were confirmed through extensive simulations, showing high training efficiency and predictive accuracy. Future work will focus on online federated optimization methods and adaptive algorithms for continuous model updating in dynamic IoT and MEC environments.

\ifCLASSOPTIONcompsoc
  \section*{Acknowledgments}
\else
  \section*{Acknowledgment}
  
\fi

  This research was fully supported by the Australian Government through the Australian Research Council’s Discovery Projects funding scheme (DP220101823).

\ifCLASSOPTIONcaptionsoff
  \newpage
\fi



%
\bibliographystyle{IEEEtran}
\bibliography{references}

%
\vspace{-5mm}
\begin{IEEEbiography}[{\includegraphics[width=1in,height=1.25in,clip,keepaspectratio]{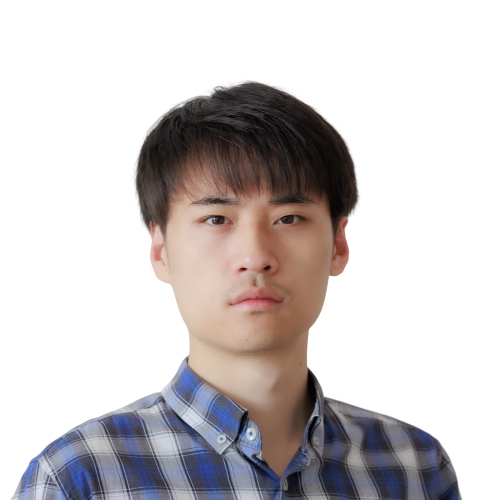}}]{Jiahui Bai} received a Bachelor of Science (Computer Science) (1st Class Honours) from RMIT University. He is currently a PhD student at RMIT University. His primary research interests include Internet of Things (IoT), machine learning, network security, and distributed systems. His publication appears in ICWS.
\end{IEEEbiography}

\begin{IEEEbiography}[{\includegraphics[width=1in,height=1.25in,clip,keepaspectratio]{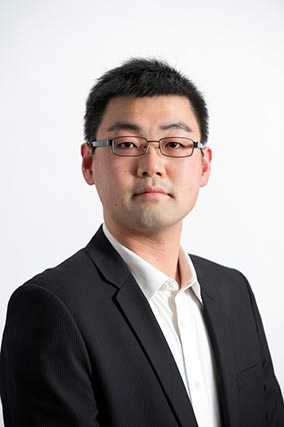}}]{Hai Dong} (Senior Member, IEEE) received the Ph.D. degree from Curtin University, Perth, Australia. He is currently a Senior Lecturer with the School of Computing Technologies, RMIT University, Melbourne, Australia. His primary research interests include services computing, edge computing, blockchain, cyber security, machine learning, and data science. 
His publications appeared in ACM Computing Surveys, IEEE Transactions on Industrial Electronics, IEEE Transactions on Industrial Informatics, IEEE Transactions on Mobile Computing, IEEE Transactions on Services Computing, IEEE Transactions on Software Engineering, ASE, ICSOC, ICWS, etc.
\end{IEEEbiography}

\begin{IEEEbiography}[{\includegraphics[width=1in,height=1.25in,clip,keepaspectratio]{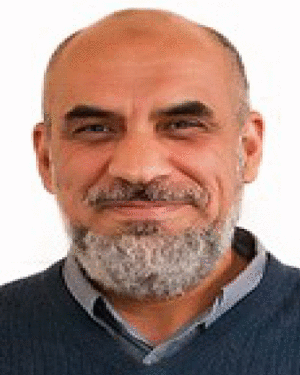}}]{Athman Bouguettaya} (Fellow, IEEE) is a Professor in the School of Computer Science at the University of Sydney, Australia. He received his PhD in Computer  Science  from  the  University  of  Colorado at  Boulder  (USA)  in  1992.  
He  was  previously Science Leader in Service Computing at CSIRO ICT Centre, Canberra. Australia. Before that, he was a tenured faculty member and Program director  in  the  Computer  Science  department  at Virginia Tech. He  is  or  has  been  on  the  editorial  boards  of  several  leading journals, including, the IEEE Transactions on Services Computing,  IEEE Transactions on Knowledge and Data Engineering, ACM Transactions on  Internet  Technology,  ACM Computing Surveys,  and VLDB Journal. He has published more than 250 books, book  chapters,  and  articles  in  journals  and  conferences  in  the  area of databases and service computing (e.g., the IEEE TKDE, the ACM TWEB,  WWW  Journal,  VLDB  Journal,  SIGMOD,  ICDE,  VLDB,  and EDBT).
He is a Fellow of the IEEE and a Distinguished Scientist of the ACM.
\end{IEEEbiography}




\end{document}